\documentclass[prl,onecolumn,superscriptaddress,nofootinbib]{revtex4-1}
\usepackage{graphicx}
\usepackage{color}
\usepackage{amsmath}
\usepackage{amsfonts}
\usepackage{mathrsfs}

\begin{document}

\title[Order in collagen fibrils]{Ordering in dense fiber bundles, the phyllotactic solution and its application to collagen fibrils}
\author{Jean Charvolin}
\affiliation{Laboratoire de Physique des Solides (CNRS-UMR 8502), B{\^a}t. 510, Universit{\'e} Paris-sud, F 91405 Orsay cedex}
\author{Jean-Fran\c cois Sadoc}
\email{sadoc@lps.u-psud.fr}
\affiliation{Laboratoire de Physique des Solides (CNRS-UMR 8502), B{\^a}t. 510, Universit{\'e} Paris-sud, F 91405 Orsay cedex}

\begin{abstract}
\emph{accepted (14 02 2013) in Biophysical Reviews and Letters (BRL)}

The shape of the cross section of a dense fiber bundle is  related to the symmetry of its molecular packing. However, this statement might be belied by type I collagen fibrils which have a rounded section of high symmetry while structural studies suggest that their molecules are assembled with a long range lateral order of lower symmetry. We examine how phyllotaxis, which is a non conventional crystallographic solution to packing efficiency in situations of high radial symmetry, can  establish a link between those two apparently conflicting points. The lateral order imposed by the algorithm of phyllotaxis, which implies an enlargement of the notion of long range lateral order beyond that used for classical crystals,
provides a basis for a new analysis of the experimental data.

\end{abstract}

\maketitle

\section{Introduction }

The organization of long molecules of biological origin in dense fiber bundles  can present various degrees of order according to the nature of the material and its environment. For instance DNA molecules can be hexagonally packed in large toroidal condensates with hexagonal cross section \cite{hud,leforestier}  while type I collagen molecules can show a high level of disorder in fibrils with circular cross section \cite{doucet}. Those two extreme examples illustrate the relation which can be expected between the local symmetry, as imposed by the packing of the molecules, and that of the aggregate, as shown by the cross section of the bundle. As a matter of fact, if the molecules are packed with a long range lateral crystalline order the cross section of their bundle is limited by the polygonal trace of reticular planes with lowest interfacial energy, if not this energy is minimized by a circular cross section.
However, this clearcut point of view has to be reconsidered as type I collagen fibrils with circular cross section
are not always as disordered as those mentioned above.
X-ray experiments indeed suggest that such fibrils  might present some long range lateral order as shown by the equatorial traces of the raw scattered intensity obtained from such fibrils \cite{hulmes81,hulmes95,bride}   where they are compared with that of a disordered fibril \cite{doucet} in figure~\ref{f1}.

These traces all present an important diffuse scattering which, after having decreased out of the beamstop, attains a broad maximum for $k \approx 5~\textrm{nm}^{-1}$. On trace (a) this diffuse scattering is smooth  while on traces (b, c) small slightly broadened peaks merge out from it in the $1<k<5~\textrm{nm}^{-1}$ range as shown by vertical arrows. The maximum at $k \approx 5~\textrm{nm}^{-1}$ and the global shape of the diffuse scattering correspond well to a highly disordered organization with a mean intermolecular distance of $1.3~\textrm{nm}$, but the peaks observed for $1<k<5~\textrm{nm}^{-1}$ suggest the presence of some kind of order with repeat distances larger than the intermolecular one. Thus these observations reveal a complex material in which disorder, the diffuse scattering, and some kind of order, the discrete bumps, can coexist without the reticular planes of the latter manifesting themselves in a facetted cross section with low symmetry. This complexity was indeed pointed out for the first time forty five years ago\cite{ramachandran} and is regularly addressed since then \cite{prockop,woodhead}.

\section{Conciliating order and disorder in type I collagen fibrils}

The data are usually analyzed considering that the formation of type I collagen fibrils proceeds along two steps as shown in figure~\ref{f2}.
The segments of triple helices are densely organized in an overlap region but in a gap region one segment of triple helix out of five is replaced by a vacancy and the distribution of the vacancies form periodic layers of Hodge-Petruska staggering.
%
%
\begin{figure}[tbp]
\includegraphics{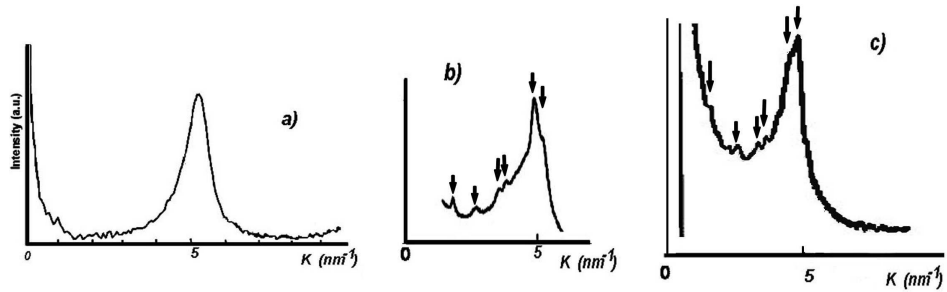}

\caption{ X-ray scatterings along the direction normal to the fibril axis, equatorial traces, obtained from native tendons by Doucet et al.(2011) (a), Hulmes et al.(1981) (b)  and  MacBride et al.(1997) (c).
We present here only raw equatorial traces of the scattered intensities. They have not been submitted to any computational treatment aiming at suppressing their background scatter in order to isolate the small discrete peaks merging out from it in b) and c). }
\label{f1}
\end{figure}
%
%
Reviews and analysis of the large number of models proposed to interpret the experimental data on the basis of this description can be found in\cite{hulmes95,wess}. Those models are of very different natures, going from a purely crystalline organization to a liquid-like one, including multiple-start spirals and concentric ring models. A rather satisfying agreement is obtained considering that :\\
-	the vacancies and the triple helix sections at the gap/overlap interface are organized according to the x,y plane of a triclinic crystal to give account of the discrete scatterings,\\
-	the gap and overlap regions are disordered, eventually more in the first,  to give account of the background scatter,\\
-	the whole fibril is described as a polycrystalline assembly of  monocrystalline grains with triangular basis separated by radial grain boundaries in order to approach a circular section by multiplying the number of monocrystalline facets at the periphery of the polycrystal.

%
%
\begin{figure}[tbp]
\includegraphics{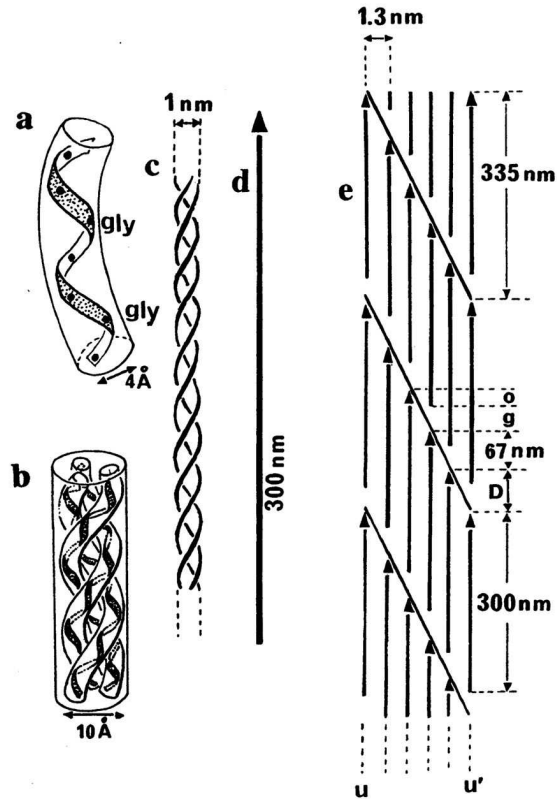}

\caption{Left-handed collagen molecules (a) assemble in right-handed triple helices (b,c) represented by rods (d) which in turn are proposed to be assembled with a regular shift (e), the so-called Hodge-Petruska  staggering, in order to create regions of gap ``g '' and overlap `` o '' to give account of the longitudinal striations of the fibrils.}
\label{f2}
\end{figure}
%
%
We recently proposed to look at this structural problem along a quite different way inspired by the iterative process of phyllotaxis, a non conventional crystallographic solution to packing efficiency in situations of high radial symmetry \cite{charvolinsadoc,sadocriviercharvolin}. This process ensures to each of the molecules the most homogeneous and isotropic local environment  possible, at the price of local metric and topological disorders which interact to build a long range self-similar organization with high radial symmetry. In\cite{charvolinsadoc} we considered parallel  triple helices in fibrils of small section, represented by a phyllotactic pattern of $2500$ points, and it  appeared that the calculated X-ray scatterings presented features which could contribute to that observed along the direction normal to the fibril axis in the $1<k<5~\textrm{nm}^{-1}$ range.

We develop here this preliminary approach considering phyllotactic patterns with larger numbers of points, up to 15704 approaching  the most common sizes of the fibrils, and making full use of our recent study of the symmetries presented by such patterns \cite{sadocriviercharvolin}. This development shows that a long range  order can be compatible with a circular cross section, an order of course different from that of classical crystallography, and that the scattered intensity expected from it can be compared with the localized scatterings observed in figure~\ref{f1}. We also introduce random displacements around the positions of these points to simulate some complementary disorder and give a better account of the background diffuse scattering observed in the same figure. Moreover, this phyllotactic order imposes an intrinsic structural complexity which suggests a new reading of the experimental data, particularly their eventual variability with the size of the cross section.

As said just above, we consider fibrils made of parallel triple helices. This model may be thought not quite realistic as, owing to the chirality of these molecules, the fibrils are most likely twisted. The pitch of the twist can be estimated to be of about $2400~\textrm{nm}$ from studies on toroidal aggregates \cite{cooper}   so that the disorientation from the center to the periphery of a common fibril having a radius of 80 nm would attain about ten degrees. This disorientation is certainly at the origin of the fan shape of the scattered intensities, but this contribution is weak enough to be ignored at this stage of the study limited to the equatorial traces. It should however not be forgotten that, although weak, such a twist might play a role in the lateral growth of fibrils as shown by a thermodynamical description of F-actin bundles \cite{grason1,grason2}  and a geometrical description of DNA toroidal aggregates \cite{charvolinsadocDNA}. It might also contribute to the mechanical properties expected from fibrils  \cite{neukirch,bozec}. In the present model we consider two contributions at the diffraction pattern: that of overlap or that of gap regions. A simple way two introduce a chirality, in an extension of this work, will be to still consider parallel molecules, but with the gap distribution rotating along the vertical axis. This will give the same equatorial traces as given here, but a realistic fan shape of the scattered intensity away from this trace.

\section{ Phyllotaxis}

The lateral organization of  parallel fibers can be represented by their projection onto a plane normal to the axis of the bundle. We build the phyllotactic organization in this plane. In the case of type I collagen two situations are to be considered corresponding respectively to the overlap regions, in which all points are occupied, and to the gap regions, in which one point over five is a vacancy.

\subsection{Phyllotactic pattern for an overlap region}

\begin{figure}[tbp]
\includegraphics{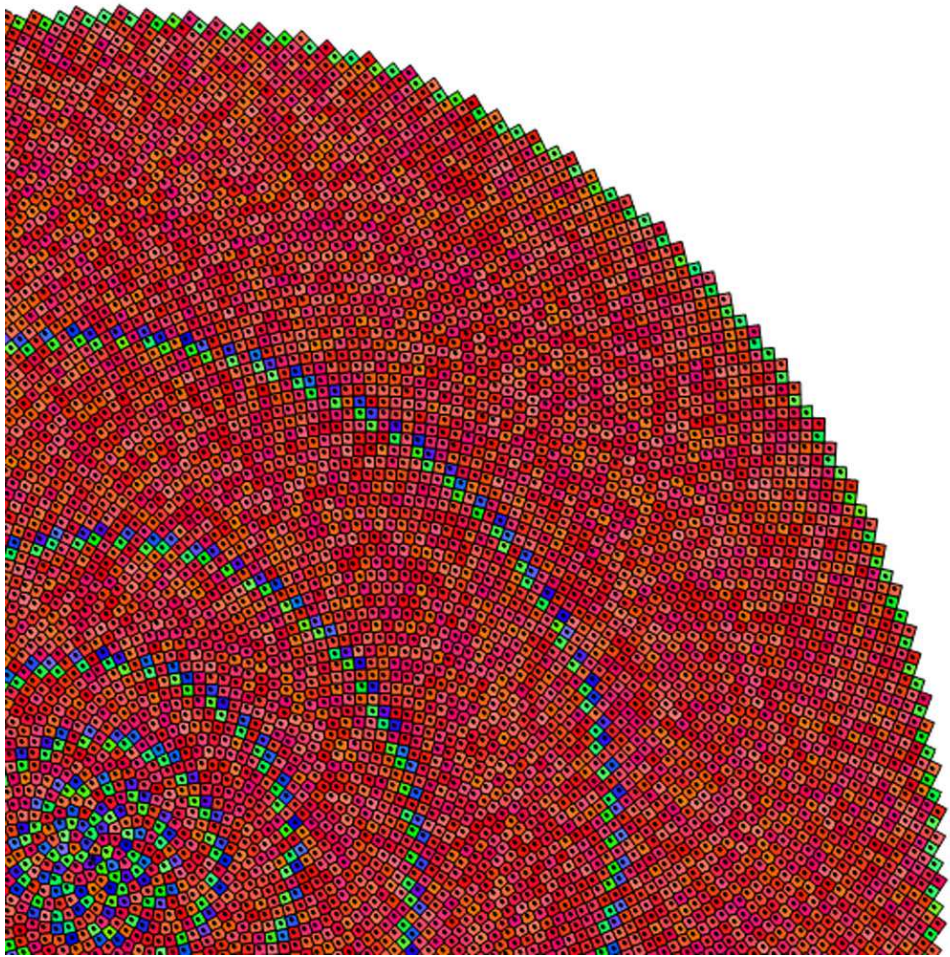}

\caption{A quadrant of a set of 15704  points organized according to the algorithm of phyllotaxis with the golden ratio.
The radius of this quadrant is about $80 \textrm{nm}$.
Each point is surrounded by its polygonal Voronoi cell whose number of sides corresponds to the number of first neighbors around this point. Blue cells are pentagons, red cells hexagons and green cells heptagons, those polygons are not necessarily regular. In the case of a point in a hexagonal Voronoi cell  three spirals, called parastichies join the first neighbors of the point. These parastichies are Fermat spirals, which are not to be confused with the Archimedean spirals of the spiral models described in Hulmes et al. (1995). Using concepts used in metallurgy and crystallography, it is possible to consider large hexagonal domains as crystalline grains (although not perfect) separated by grain boundaries formed by defects (pentagon-heptagon dipoles). See table~\ref{tab1} for the labeling of these grains.}
\label{f3}
\end{figure}

A phyllotactic organization of points indexed by integer $s$ is built by an algorithm such that the position of point $s$ is given by its polar coordinates $r=a \sqrt{s}$ and  $\theta=2\pi\lambda s$  that is $r=a\sqrt{\theta}/ \sqrt{2 \pi \lambda}$   which is the equation of a Fermat spiral here called the generative spiral. The area of the circle of radius $r$ which contains $s$ points is $\pi a^2 s$ so that the area per point has the value $\pi a^2$, indeed it oscillates close to this value for small $s$ then converges towards it\cite{sadocriviercharvolin}. The most homogeneous and isotropic environment, or the best packing efficiency in radial symmetry, is obtained with  $ \lambda=1/\tau$  where  $\tau$ is the irrational golden ratio $(1+ \sqrt{5})/2$ \cite{jean1,jean2,ridley}. A sector of a phyllotactic pattern for $N=15704$  points with their Voronoi cells is shown in figure~\ref{f3}.

%
\begin{figure}[tbp]
\includegraphics{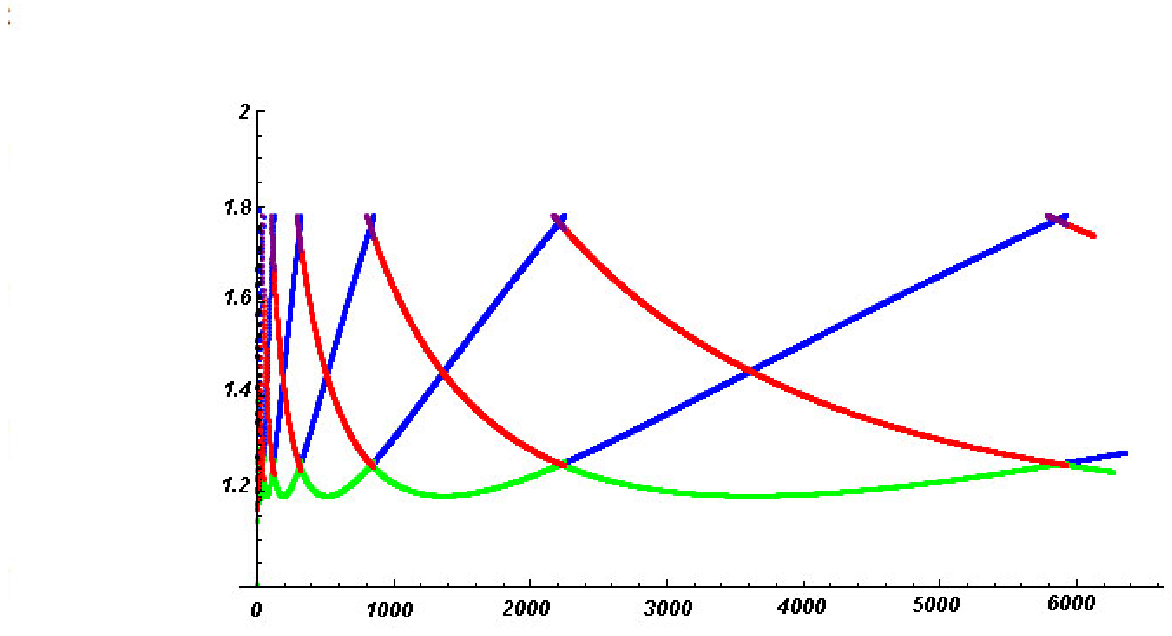}

\caption{Distances between first neighbor points with the parameter $a=0.71$ leading to a mean distance close to $1.3~\textrm{nm}$ as in collagen. The three tones correspond to the three parastichies, the upper and lower crossings on the same verticals correspond to grain boundaries and the intermediate ones to the cores of hexagonal grains (see table~\ref{tab1} for grains labeling). All distances are confined in the same domain  whatever the size of the structure from  1.18 nm to 1.77 nm.
}
\label{f4}
\end{figure}

%
%
%
%
In such a pattern, pentagons and heptagons are topological defects distributed among the hexagons. They appear concentrated in narrow circular rings with constant width which separate large rings of hexagons whose width increases as one moves from the core towards the periphery. In the narrow rings, pentagons and heptagons are associated in dipoles separated by hexagons whose shape is close to that of a square with two corners cut. The rings of dipoles are indeed grain boundaries separating hexagonal grains\footnote{The grain boundaries are here circular while they are radial in the polycrystalline model of  Hulmes et al. (1995)\cite{hulmes95}, a dense organization with radial symmetry can not be built without defects. }. The underlying arithmetic of this organization is the Fibonnacci series, $f_u=f_{u-1}+f_{u-2}$ from $f_0=0$ and $f_1=1$, which makes the organization self-similar as it is invariant by a change of scale  $\tau^n$ as summarized in table~\ref{tab1}(see\cite{sadocriviercharvolin}).
This is also made apparent considering the evolutions of the distances between first neighbors shown in figure~\ref{f4}.

\begin{table}
\caption{\label{tab1} Cell types in the successive rings. The number $s$ is that of each point on the generative spiral, the first neighbors of point $s$ have numbers $s+\delta s$ where $\delta s$ are Fibonnacci numbers, all the same for a given ring except in the core of the pattern. The large hexagonal rings or grains are labeled by the rank $u$ of the Fibonnacci number $f_u$ corresponding to the medium value of $\delta s$ in a ring, the $\delta s$ are also the numbers of parastichies of each type in the ring. The narrow rings or grain boundaries are marked by   $\|$. }
\begin{tabular*}{\textwidth}{@{}l*{15}{@{\extracolsep{0pt plus
12pt}}l}}
\toprule
$u$&Cell type&number of cells&$s$ from &to&neighbor separations  $\delta s$\\
\hline
&pentagon&2&0&1&1,2,3,4,5 or -1,2,3,5,8\\
&hexagon&1&2&2&-2,2,3,5,8,13\\
&heptagon&3&3&5&(-3,-2,2) or (-4,-3,-2) or (-5,-3,-2),3,5,8,13\\
&hexagon&1&6&6&-5,-3,3,5,8,13\\
&pentagon&2&7&8&-5,-3,5,8,13\\
&hexagon&1&9&9&-8,-5,-3,5,8,13\\
$\|$&hexagon&5&10&14&-8,-5,5,8,13,21\\
$\|$&heptagon&3&15&17&-13,-8,-5,5,8,13,21\\
$\|$&hexagon&5&18&22&-13,-8,-5,8,13,21\\
$\|$&pentagon&8&23&30&-13,-8,8,13,21\\
7&hexagon&2&31&32&-21,-13,-8,8,13,21\\
$\|$&heptagon&13&33&45&-21,-13,-8,8,13,21,34\\
$\|$&hexagon&8&46&53&-21,-13,-8,13,21,34\\
$\|$&pentagon&13&54&66&-21,-13,13,21,34\\
8&hexagon&34&67&100&-34,-21,-13,13,21,34\\
$\|$&heptagon&21&101&121&-34,-21,-13,13,21,34,55\\
$\|$&hexagon&13&122&134&-34,-21,-13,21,34,55\\
$\|$&pentagon&21&135&155&-34,-21,21,34,55\\
9&hexagon&134&156&289&-55,-34,-21,21,34,55\\
$\|$&heptagon&34&290&323&-55,-34,-21,21,34,55,89\\
$\|$&hexagon&21&324&344&-55,-34,-21,34,55,89\\
$\|$&pentagon&34&345&378&-55,-34,34,55,89\\
10&hexagon&422&379&800&-89,-55,-34,34,55,89\\
$\|$&heptagon&55&801&855&-89,-55,-34,34,55,89,144\\
$\|$&hexagon&34&856&889&-89,-55,-34,55,89,144\\
$\|$&pentagon&55&890&944&-89,-55,55,89,144\\
11&hexagon&1221&945&2165&-144,-89,-55,55,89,144\\
$\|$&heptagon&89&2166&2254&-144,-89,-55,55,89,144,233\\
$\|$&hexagon&55&2255&2309&-144,-89,-55,89,144,233\\
$\|$&pentagon&89&2310&2398&-144,-89,89,144,233\\
12&hexagon&3384&2399&5782&-233,-144,-89,89,144,233\\
$\|$&heptagon&144&5783&5926&-233,-144,-89,89,144,233,377\\
$\|$&hexagon&89&5927&6015&-233,-144,-89,144,233,377\\
$\|$&pentagon&144&6016&6159&-233,-144,144,233,377\\
13&hexagon&9167&6160&15326&-377,-233,-144,144,233,377\\
\hline
\end{tabular*}
\end{table}

%
%
%
%
%

\subsection{Phyllotactic pattern for a gap region}

\begin{figure}[tbp]
\includegraphics{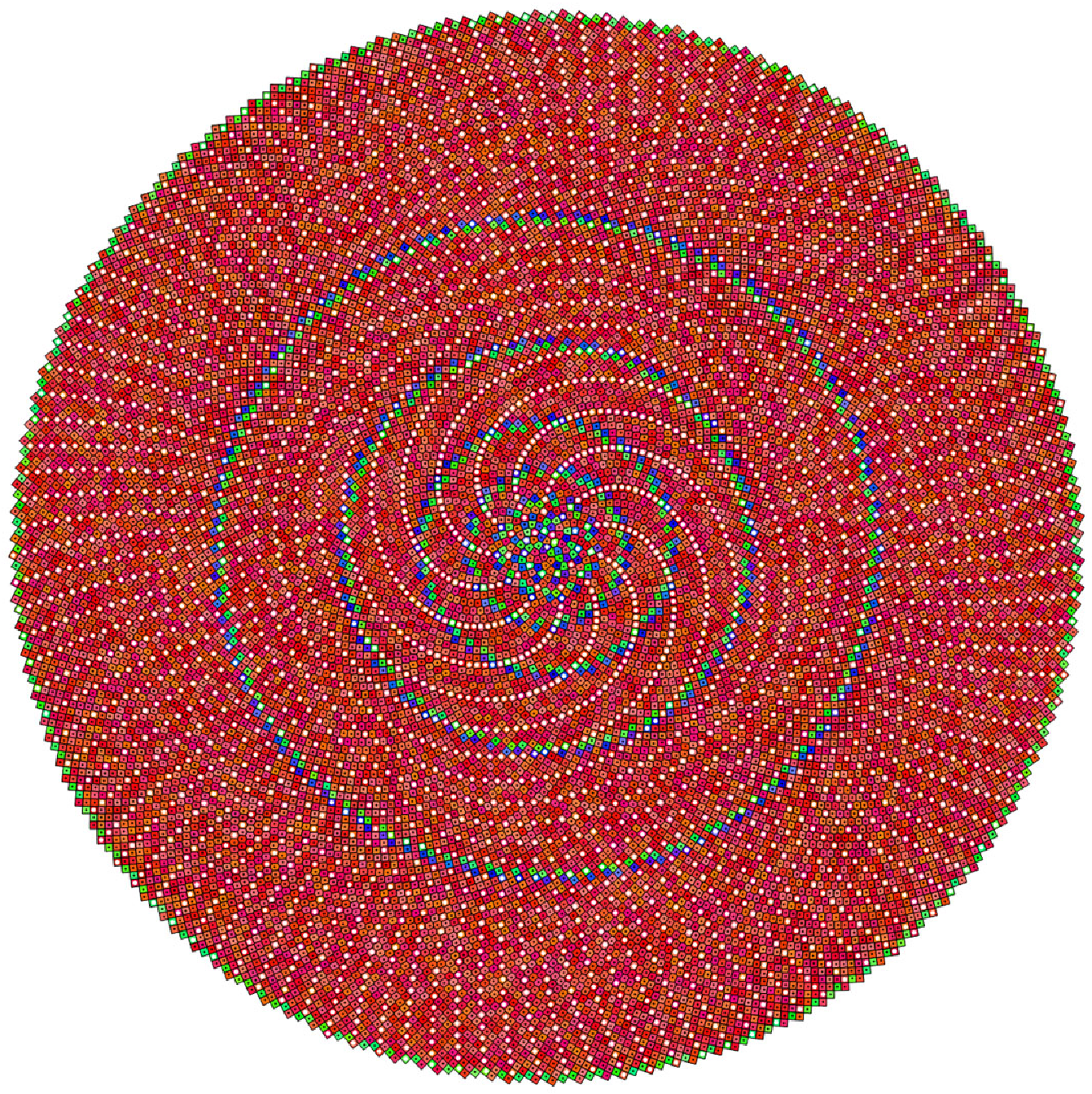}

\caption{ A  set of 15704  points organized according to the algorithm of phyllotaxis with the golden ratio and in which one point out of five is replaced by a vacancy, a white spot, in the curse of the development of the generative spiral.}
\label{f5}
\end{figure}

The phyllotactic pattern of a gap is similar to that of an overlap except for the fact that one point out of five must be replaced by a vacancy. The very nature of the iterative process leads to do this replacement all along the generative spiral. Then, the vacancies appear distributed on new spirals in a manner respecting the periodicity required by the staggering of the triple helices, as shown in figure~\ref{f5}.
%
%
%
%
%
%
%
%
%
\begin{figure}[tbp]
\includegraphics{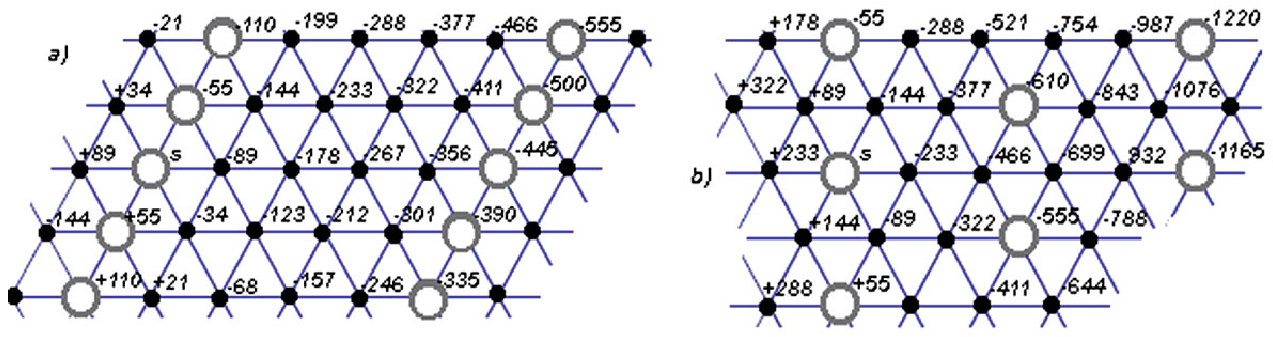}

\caption{ Numberings of the points surrounding a point $s$ in grain $u=10$ (see table~\ref{tab1}) with one  $\delta s=f_u$  multiple of 5 (a) and grain $u=12$ without  $\delta s$  multiple of 5 (b), for simpleness the Voronoi cells of the grains are here assumed to be regular hexagons and the parastiches are therefore straight lines.}
\label{f6}
\end{figure}
%
%
On these new spirals, the vacancies can be first neighbors or not. They are first neighbors in the hexagonal grains  having one $\delta s$ multiple of five but not in all others. This organization can be built from table~\ref{tab1} starting from one point $s$ in a hexagonal grain and using the $\delta s$ of this grain to number its surrounding sites step by step. Two examples are drawn in figure~\ref{f6} based on regular hexagons for a point $s$ in grain   $u=10$, where  $|\delta s| =89,55,34$ with one  $\delta s$  multiple of $5$, and a point $s$ in grain $u=12$ where  $|\delta s | = 233,144,89$ without  $\delta s$  multiple of $5$.
If the starting point $s$ of figure~\ref{f6} is considered as a vacancy and its number $s$ is a multiple of 5, the vacancies are first neighbors in grain 10, second neighbors in grain 12 and build lattices whose elementary cells are shown in figure~\ref{f7} in these drawing based on regular hexagons.
%
%
%
\begin{figure}[tbp]
\includegraphics{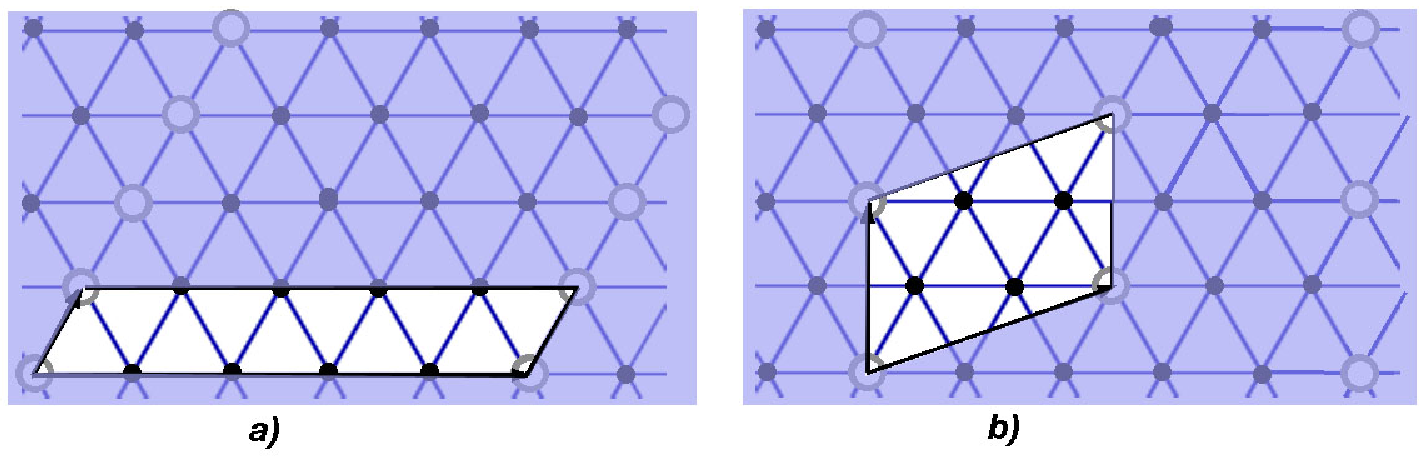}

\caption{ Elementary cells determined by the vacancies, open circles, in the triangular network of points of figure~\ref{f6} for grain 10 with one  $\delta s$  multiple of 5 (a) and grain 12 without  $\delta s$  multiple of 5 (b). The multiplicity of both cells in the hexagonal lattice is 5.}
\label{f7}
\end{figure}
%
%

In each case, (a) or (b), the staggering of the triple helices shown in figure~\ref{f2}e can be lined up along straight lines joining first neighbors in agreement with what is observed along the spirals of the parastichies in the pattern of figure~\ref{f5}. Of course, in a grain of the phyllotactic pattern, the Voronoi cells are hexagonal in the topological sense only and the elementary cells (a) and (b) drawn in figure~\ref{f7} suffer deformations when moving in between the grain boundaries limiting the grain. The notion of elementary cell is no longer that used in classical crystallography. It has lost its metric content, distances and angles defining the positions of the sites, to keep only its topological part, neighboring relations.

As shown in table~\ref{tab2}, those elementary cells (a) or (b) are alternatively localized in concentric domains of increasing sizes corresponding to groups of three consecutive grains, domains A, and groups of two consecutive grains, domains B.
As in domains A one number  $\delta s$   of parastichies is a multiple of 5 the simple elementary cell (a) can be periodically packed in the circular grains and the vacancies are first neighbors. This is no longer possible in domains B where the numbers of parastichies  $\delta s$ are not multiple of 5 and impose the more complex elementary cell (b).

\begin{table}
\caption{\label{tab2}  Grains from $u=9$ to $u=21$,  $\delta s$  in those grains, domains A (in which one of the  three  $\delta s$ is multiple of 5, in bold face) and B (no $\delta s$ multiple of 5) occupied by elementary cells (figures~\ref{f7}-a and \ref{f7}-b), sizes of the domains and their numbers of points. See table~\ref{tab1}  for the grain labeling. }
\begin{tabular*}{1\textwidth}{@{\extracolsep{\fill}}|c|c c|c|c|c|c|c|}
\toprule
Hexag.&$s$ from & to &$\delta s$ &Concentric&inner&external& Number \\
 grains &    &   &  &domains &radius (nm)&radius &of points\\
\hline
9&156&289&$\bf{55}$,34,21 & A  &8.5& & \\
10&379&800&89,$\bf{55}$,34 & A & & &2010\\
11&945&2165&144,89,$\bf{55}$ & A  & &31.7& \\
\hline
12&2399&5782&233,144,89 & B  &33.43 & & \\
13&6160&15326&377,233,144 & B & &84.25 &12928\\
\hline
14&15937&40428&$\bf{610}$,377,33 & A  &86.16 &   &  \\
15&41416&106335&987,$\bf{610}$,377 & A &  &  & 263250\\
16&107933&279186&1597,987,$\bf{610}$ & A  &  &360.6&   \\
\hline
17&281771&732208&2584,1597,987 & B  &362.3&  &  \\
18&736390&1919036&4181,2584,1587 & B &  & 945.5&1637266\\
\hline
19&1925822&5027483&$\bf{6765}$,4184,2584 & A  &947.2 & & \\
20&5038430&13167594&10946,$\bf{6765}$,4184 & A  & & & \\
21&13185306&34510721 &17711,10946,$\bf{6765}$ & A  & &4009.8 &32584899 \\
\hline
\end{tabular*}
\end{table}

In a domain A such as that of grains $9, 10, 11$ where $\delta s=55$  there are $55/5=11$ spirals of first neighbor vacancies. These spirals remain in the domain B of grains $12, 13$, but  vacancies are now second neighbors and the elementary cell is of type (b). In this domain  the $11$ spirals are aligned along edges of the elementary cells. In the next domain A of grains $14, 15, 16$ the $610/5=122$ spirals would appear formed with first neighbor  vacancies. Nevertheless they are discernable  before in grains $12, 13$ formed with second neighbors. These spirals are aligned along edges of the (b) cells. So in these grains one family of cell edges form $11$ spirals and the other family form $122$ spirals.

\section{Phyllotaxis vs classical crystallography}

In a phyllotactic pattern, most of the points have six first neighbors, as expected for a dense system, but those first neighbors can not draw regular hexagons. They are distorted like in a flat 2D crystal bent in its plane so that fluctuations of distances and angles introduce a metric disorder. Moreover, a few points have five or seven first neighbors which are defects, disclinations, introducing a topological disorder. Those two disorders interact so as to build a concentric organization of grains of points with six first neighbors separated by grain boundaries in which the topological defects are associated in dipoles equivalent to dislocations. Those dislocations are distributed according to an inflation/deflation quasicristalline rule along the grain boundaries whose radii vary as the Fibonacci sequence. This organization is therefore not invariant by the translation and rotation operations of classical crystallography, but is self-similar. The grain boundaries which, in ordinary materials, are extrinsic defects imposed by deformations of the material are here intrinsic. In the grains, the six first neighbor points of each point are aligned along the three families parastichies which can be seen as equivalent to the reticular planes of classical crystallography. Crossing a grain boundary, from one grain to the next, the dislocations present in it introduce the new parastichies, hence the new points, needed to maintain the density constant (see\cite{sadocriviercharvolin} for remarks in this paragraph).

The  staggering of the triple helices introduces vacancies which are organized along new spirals in each gap where they can be either first neighbors or not. Each of these two distributions can be said ordered, but once again not in the sense of classical crystallography, as this order concerns only the neighboring relations of one vacancy among four cross sections of triple helices. In one case, elementary cell (b), the disposition of the four cross sections of triple helices around one vacancy is very close to that existing in the x,y plane of the triclinic cell proposed in some previous models\cite{hulmes95,wess}. However the fluctuations of distances and angles characteristic of phyllotaxis can not let expect Bragg peaks in the X-ray diagrams.
Finally, those two distributions of vacancies are themselves organized in alternated concentric domains whose sizes increase with the radius of the pattern, in a manner related to the Fibonacci series. This is an intrinsic structural heterogeneity which has no equivalent in classical crystallography.

\section{Scattered intensities expected from phyllotactic  patterns}

\begin{figure}[tpb]
\caption{Intensities scattered by overlaps, taking account of a  form factor corresponding to a molecular diameter  of 1~nm,   for N=2500 (a), N=6500 (b) and N=15704 (c). }
\includegraphics{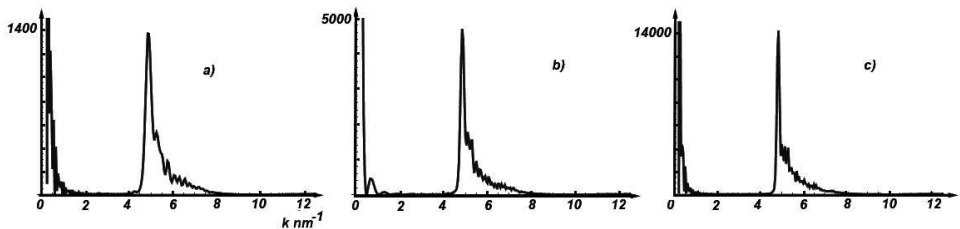}

\label{f8}
\end{figure}

\begin{figure}[tpb]
\caption{Intensities scattered by gaps, taking account of a  form factor corresponding to a molecular diameter  of 1~nm, for N=2500 (a), N=6500 (b) and N=15704 (c).}
\includegraphics{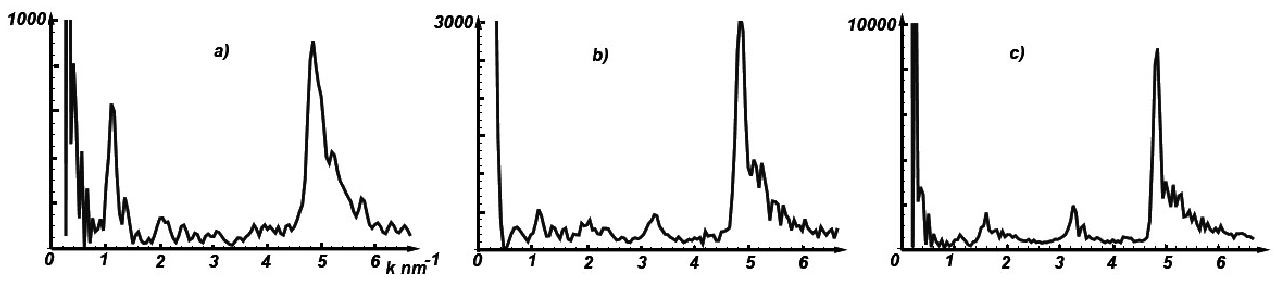}
\label{f9}
\end{figure}

The main information concerning the lateral organization of collagen fibrils being provided by X-rays scattering experiments, we calculate the scatterings expected from the above phyllotactic templates for comparison. As the triple helices are normal to the patterns, the equatorial traces of the intensities scattered by the cylindrical overlaps and gaps containing them are those of their respective patterns. The scattered intensity expected from a distribution of points is the square of the amplitude of its Fourier transform, but, as the points are representative of triple helices having a diameter of $1~\textrm{nm}$, the points are to be changed into disks and the intensity decreases as the form factor of the disk. The results are shown in figures~\ref{f8} and~\ref{f9} for the overlap and gap patterns respectively, in both cases for patterns of increasing sizes.

 %
%
%

%
%

%
%
%
The intensity scattered by overlaps show that, whatever the size of the pattern, the scattering is concentrated in the two regions $k<1~\textrm{nm}^{-1}$ and $k\approx5~\textrm{nm}^{-1}$. The scattering in the small wave vectors region corresponds to long range fluctuations such as the size of the pattern and the circles of dipoles. The scattering around $5~\textrm{nm}^{-1}$ is to be related to the local organization of the triple helices.
In the intensity scattered by the gaps, the features of the scatterings produced by the overlaps are still present  but new ones appear at $k\simeq1~\textrm{nm}^{-1}$ and in the range $1<k<5~\textrm{nm}^{-1}$, the latter, hardly discernable for N=2500, become more important for $N=6500$ then quite obvious for $N=15704$. These are not strictly Bragg peaks, but peaks with a finite width and  broadened  feet  which contribute to a raising of the baseline.

The scatterings observed at $k\simeq1~\textrm{nm}^{-1}$, one fifth of the position of that observed at $k\simeq5~\textrm{nm}^{-1}$, is that expected from elementary cells (a) and those observed in the $1<k\leq5~\textrm{nm}^{-1}$ range  are those expected from elementary cells (b). The evolutions of these scatterings with the sizes of the patterns can be understood from table~\ref{tab2}.  The patterns with $N=2500$ is dominated by grains 9 to 11, i.e. a domain A with the elementary cell (a), increasing the size of the pattern to $N=6500$ then $N=15704$ adds grains $12$ then $13$, both included in a domain B with elementary cell (b)  whose contribution  increases accordingly. If N was increased beyond 15704 a domain A would be added whose contribution would become rapidly dominant as it contains about 20 times more points than the two first domains.

\section{Comparison with experiments}

\subsection{X-rays data}

\begin{table}
\caption{\label{tab3} Positions measured on raw equatorial traces (A) Hulmes 1981,1995, Mc Bride 1997 , on traces from which the background scatter has been subtracted (B) Wess 1998, Orgel 2001  and on the calculated trace (C) of figure 9c.}
\begin{tabular}{|l|c|c|c|c|c|c|c|c|r|}
 \hline
& \multicolumn{9}{c|} { peaks ($\textrm{nm}^{-1}$)}\\
\hline
$\alpha$ &Hulmes et al. (1981)\cite{hulmes81} & & 1.65&2.5&3.3&3.59&4.6&4.98& \\
\cline{2-10}
&Hulmes et al. (1995)\cite{hulmes95}& & 1.62&2.65&3.27&3.54&4.5&4.79& \\
\cline{2-10}
 &McBride (1997)\cite{bride} & & 1.64&2.5&3.28&3.57&4.58&4.92& \\
 \hline
$\beta$ &Wess et al. (1998)\cite{wess} & & 1.4&2.1;2.3&3.19&3.47&4.67&5.17&(5.7) \\
 \cline{2-10}
       &Orgel et al.(2001)\cite{orgel} & & 1.3&2.2;2.3&3.2&3.47&4.67&5.04&(5.7) \\
 \hline
 $\gamma$ &Phyllotaxis &1.05 & 1.6&2.4&3.2&3.4&4.5&4.95&(5.6) \\
 \hline
\end{tabular}
\end{table}

Raw equatorial traces obtained with native samples have been already shown in figure~\ref{f1} and described in the introduction. The positions of the peaks merging out from the the background scatter in figures 1b,c reasonably agree with those obtained for a gap represented by a phyllotactic pattern of N=15704 points, as shown in table~\ref{tab3}.
Such an agreement is not surprising as the distribution of the vacancies in the domains B of the gaps is, at least as far as its topology is concerned, close to that in the crystalline cells proposed in previous works. The widths of these discrete scatterings results from the metric distortions of the Voronoi cells and their distribution within the frame of the self-similar reproduction. The scattering of domain A at $k=1~\textrm{nm}^{-1}$ has not been mentioned up to now. A trace of it might be found in figure~\ref{f1}a from ref.\cite{doucet} but may be lost in
the intense small angle scattering observed around the beamstops in the other references.

\begin{figure}[tbp]
\caption{Scattered intensity obtained from a pattern having the size of a pattern with N=15704 points but in which the only scattering centers are the points corresponding to the vacancies.}
\includegraphics{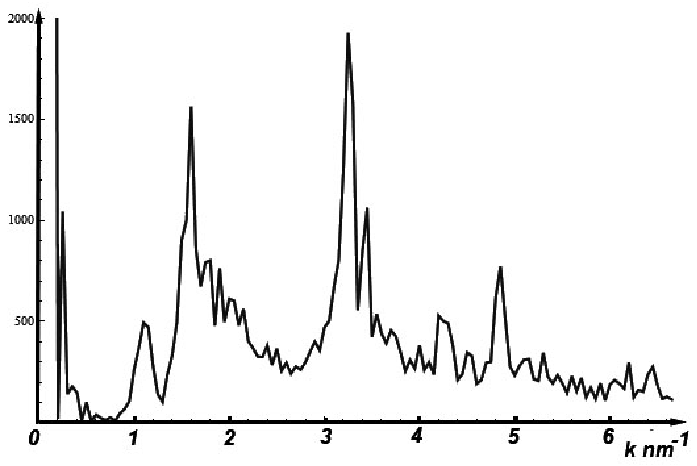}
\label{f10}
\end{figure}

When heavy atoms are incorporated in the sample in order to enhance its contrast the general shape of the scattered intensity changes with respect to that of native samples\cite{hulmes81}. Essentially the very intense scattering at $k\approx 5~\textrm{nm}^{-1}$ decreases and that at $k\approx 3~\textrm{nm}^{-1}$ increases. This change might result from a change of structure induced by the incorporation process, but it also might be that the heavy atoms are predominantly localized in the vacancies of the gaps enhancing the weight of the scattering expected from those vacancies as shown on figure~\ref{f10}.

  %
%
%
%
%

\begin{figure}[tbp]
\caption{ Scattered Intensities by N=15704 sites without disorder (a), with an increasing disorder (b,c). The histogram of radial displacements is indicated (azimuthal displacements are similar). These calculated intensities take account of  an equal contribution from overlaps and gaps.}
\includegraphics{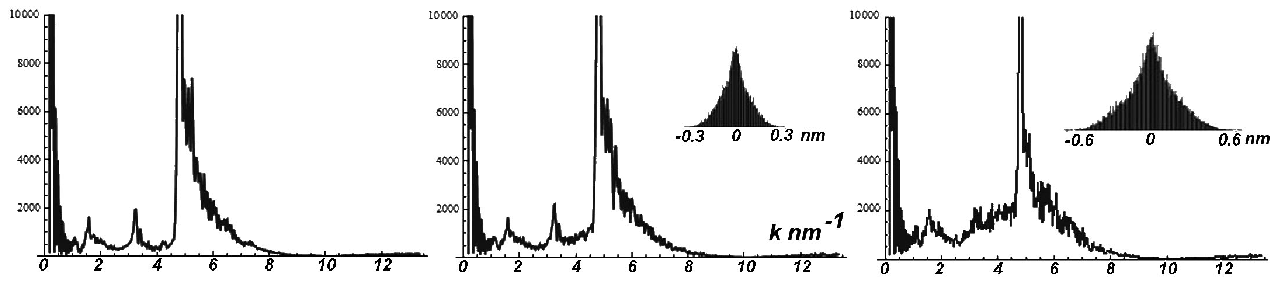}
\label{f11}
\end{figure}

But, in the experimental data, those discrete scatterings contribute for no more than 5 to 10 percent of the total intensity \cite{prockop} which is dominated by a background scatter bearing evidence for  an important level of disorder in the sample. Certainly, a phyllotactic organization contains an intrinsic metric disorder associated to the deformations of the Voronoi cells, but as shown by the calculated scattered intensities presented in figures~\ref{f8} and~\ref{f9}, these fluctuations are not sufficient to give account of the intense background scatter observed. Indeed, phyllotaxis is but a geometrical template and a disorder must be introduced. Such a disorder can be simulated \footnote{ We do not establish a distinction between gaps and overlaps at this stage although the disorder might be more important in the first because of the vacancies.} making the position   of each point $s$ fluctuates  randomly\cite{rivier84,rivier92}  with an amplitude characterized by the histograms of figure~\ref{f11} where the evolution of the calculated spectra as this disorder increases is shown.
  %
%
%
%
%

A randomly distributed disorder does not change the shape of the scattering, it just change the amplitude and adds a constant background. It is analogous to what happens when such   disorder is introduced in a crystal: the width of Bragg peaks  remain very small and a Laue continuous background appears, but if correlations are introduced in the disorder they lead to a broadening of localized peaks.  To introduce short range correlations we have convoluted displacement with a Gaussian function having a width close to the distance between second neighbors. This method  introduces short range correlations in the disorder between first and second neighbor points but not long distance ones, nevertheless the broadlines of the experimental spectra are rather well reproduced: small discrete bumps are superposed on an background increasing with $k$ and culminating at $k\approx 5~\textrm{nm}^{-1}$. The positions of the small discrete peaks are not affected by the disorder, only their width, this holds to the fact that the displacements on short distances are small compared to the characteristic distances between rows of vacancies.

\subsection{Electron microscopy}

Micrographs of ultrathin sections of rat tail tendon prepared to provide optimum preservation of the native fibril structure with sufficient contrast for electron microscopy give an impression of tangentially oriented lines of density which may be interpreted as concentric circles or spirals\cite{hulmes81}. The study by optical diffraction with an equivalent aperture of $120-180~\textrm{nm}$ of micrographs of fibril sections with large radii of   210 nm measures a separation period of   $4.6~\textrm{nm}$ and indicates that this periodicity is detected predominantly towards the edge rather than the center of the fibrils. This can be put in correspondence with the heterogeneous organization of a phyllotactic pattern of similar size as deduced from table~\ref{tab2}. In such a pattern a peripheral ring of thickness $120~\textrm{nm}$ made of a domain A surrounds a central core of radius $80~\textrm{nm}$ made of a domain B predominantly. In the peripheral domain A first neighbor vacancies outline 122 spirals distant by $3$ to $5~\textrm{nm}$ and arriving at a small angle on the limit of the pattern while in the central domain B second neighbor vacancies outline a more complex combination of spirals.

\section{Conclusion}

Electron microscopy studies show that type I collagen fibrils have a rounded cross section while some X-rays scattering studies suggest that their components, the so-called triple helices, can be assembled with a long range lateral order. This order cannot be crystalline as the topological constraint of circular symmetry would not be compatible with its long range propagation. The most homogeneous and isotropic dense organization compatible with circular symmetry is indeed the self similar one issued from the iterative process of phyllotaxis. We considered here that the circular symmetry is imposed to the assembly of triple helices, for instance by the interfacial tension of the collagen fibril, and that this assembly adapts itself to this constraint in order to achieve the best packing efficiency typical of phyllotaxis. This adaptation while the fibril grows requires a certain internal mobility which is indeed suggested by NMR studies\cite{jelinsky}.

When a long range lateral order is detected by X-rays scattering studies, it is characterized by distances larger than the mean distance between triple helices. This is most likely to be associated with the fact that, according to the Hodge-Petruska staggering, one section of triple helix out of five in the gaps of the fibrils is replaced by a vacancy. Doing this replacement all along the generative spiral of phyllotaxis leads to a distribution of vacancies in the phyllotactic pattern of the gaps such that the staggering periodicity is established along parastichies of the pattern following an organization in concentric alternated domains presenting two different structures.

The scattered intensity expected from one of these structures is made of discrete scatterings whose positions compare well with those observed. The distribution of vacancies is indeed, at least as far as its topology is concerned, close to that in the crystalline cells proposed in previous works to analyze the data. However, here, although narrow in their most intense part, these scatterings are not Bragg peaks, they have a more complex shape due to the metric fluctuations inherent to phyllotaxis. The presence of domains with another structure has not been mentioned up to now, except may be suggested by electron microscopy observations. The characteristic component of the  intensity scattered by this structure should be observed in the small angle region and might contribute to the very intense small angle scattering observed close to the beamstop in most of the observations. But also, as the widths of these domains increase rapidly when moving away from the center, the scattering of a fibril should be dominated by that of one or the other of the two domains according to its size. This could underlie the differences between the spectra shown in figure~\ref{f1}. Unfortunately, scattering studies do not give a direct access to the size of the fibrils and the samples are tendons which often contain fibrils of different sizes.

Phyllotaxis is but a zero temperature model, some thermal disorder must be introduced in the gaps as well as in the overlaps. Our simulation, although limited at local random displacements without long range correlations, leads to a rather good reproduction of the broadlines of the observed spectra. As those displacements are small compared to the characteristic distances between rows of vacancies, only the width of the scatterings expected from them is affected, not their position. This broadening increases with the disorder making these discrete scatterings less discernable, an evolution which might also contribute to the differences between the spectra shown in figure~\ref{f1}.

Finally, as suggested by the title of this article, we believe that this phyllotactic model might be useful to describe the lateral organization  of other dense biopolymer bundles such as F-actin, microtubules, cuticular collagen or some DNA condensates. However, in these cases, we can not expect to have access to the most informative data used to test this model : the scattering provided by the distribution of vacancies in the gap regions of type I collagen fibrils. Indeed, as gap regions are not seen in the biopolymer bundles quoted above, the equatorial traces of their scattered intensities should be similar to that calculated for the overlap regions of collagen which does not give any information about the medium and long range organization of the molecules. One can but rely upon electron microscopy studies, but they require the development of complex preparation procedures preserving the structure of native samples.


\section*{References}

\begin{thebibliography}{10}


\bibitem{hud}N. V. Hud and K. H. Downing,  ``Cryoelectron microscopy of $\lambda$ phage DNA 	
condensates in vitreous ice: The fine 	structure of DNA toroids'', 	
\textit{Proc. Natl. Acad. Sci. U.S.A.} \textbf{98}, $14925-14930$ (2001).

\bibitem{leforestier} A. Leforestier and F. Livolant, ``The Bacteriophage genome undergoes a succession of intracapsid phase transitions upon DNA ejection'',  \textit{J. Mol. Biol.}  \textbf{396},  $384-395$ (2010).

\bibitem{doucet} J. Doucet, F. Briki, A. Gourrier, C. Pichon, L. Gumez, S. Bensamoun and J-F. Sadoc, ``Modeling the lateral organization of collagen molecules in fibrils using the paracrystal concept'',
\textit{Jour. of Struc. Biology} \textbf{173},  $197-201$ (2011).

\bibitem{hulmes81}  D. J. S. Hulmes , J-C. Jesior, A. Miller, C. Berthet-Colominas and
C. Wolff,   ``Electron microscopy shows periodic structure in collagen fibril
cross sections'', \textit{ Proc. Nati. Acad. Sci. USA}  \textbf{78},  $3567-3571$ (1981).

\bibitem{hulmes95} D. J. S. Hulmes, T. J. Wess, D. J. Prockop, and P. Fratzl,
 ``Radial packing, order, and disorder in collagen fibrils'', \textit{Biophysical Journal}
\textbf{68}, $1661-1670$ (1995).

\bibitem{bride} McBride, D. J., Choe, V., Shapiro, J. R. and B. Brodsky,
(1997) ``Altered collagen structure in mouse tail tendon lacking the a2(I) chain'', \emph{ J. Mol. Biol.} \textbf{270}, $275–284$

\bibitem{ramachandran} G. N. Ramachandran, ``Structure of collagen at the molecular
level'', in G. N. Ramachandran (Ed.), \textit{Treatise on Collagen:
Chemistry of Collagen}  Academic Press New York, \textbf{1}, $103-184$ (1967).


\bibitem{prockop} D. J. Prockop and A. Fertala, ``The Collagen Fibril: The Almost Crystalline  Structure'',
\textit{Jour. of Struc. Biology} \textbf{122}, $111-118$ (1998).


\bibitem{woodhead}  J. Woodhead-Galloway and P. A. Machin, ``Modern theories of liquids and diffuse equatorial X-ray scattering from collagen'', \textit{ Acta Cryst. A} \textbf{32}, $368-372$ (1976).


\bibitem{wess}T. J. Wess,  A. P. Hammersley, L. Wess and A. Miller, ``A Consensus Model for Molecular Packing of Type I Collagen'', \textit{Jour. of Struc. Biology}  \textbf{122}, $92-100$ (1998).

\bibitem{charvolinsadoc}  J. Charvolin  and J-F. Sadoc , ``A Phyllotactic Approach to the Structure of  Collagen Fibrils'' \emph{Biophysical Reviews and Letters}   \textbf{6}, $13-27$ (2011).

\bibitem{sadocriviercharvolin} J-F. Sadoc , N. Rivier, J. Charvolin  and  Sadoc J-F., ``Phyllotaxis: a non-conventional crystalline solution to packing effiency in situation with radial symmetry'' \emph{Acta Cryst. A} \textbf{68}, $470-483$ (2012).

\bibitem{cooper} A. Cooper, ``The precipitation of toroidal collagen fibrils'', \emph{Biochem. J.} \textbf{112}, $515-519$ (1969).

\bibitem{grason1}  G. M. Grason, ``Braided bundles and compact coils: The structure and thermodynamics of hexagonally packed chiral filament assemblies'', \emph{Phys. Rev. E} \textbf{79}, 041919 (2009).

\bibitem{grason2}     G. M. Grason , `` Topological Defects in Twisted Bundles of Two-Dimensionally Ordered Filaments'',
    \emph{Phys. Rev. Lett.} \textbf{105}, 045502 (2010).

\bibitem{charvolinsadocDNA} J. Charvolin and J.-F. Sadoc, `` A geometrical template for toroidal aggregates of chiral macromolecules '', \emph{Eur. Phys. J. E} \textbf{25}, $335-341$ (2008).

\bibitem{neukirch}    S. Neukirch, A. Goriely and A. C. Hausrath, `` Chirality of coiled-coils : elasticity matters'',
    \emph{Phys. Rev. Lett.} \textbf{100}, 038105  (2008).

\bibitem{bozec}    M. P. E. Wenger, L. Bozec, M. A. Horton and P. Mesquida, `` Mechanical properties of collagen fibrils '', \emph{Biophysical Journal} \textbf{93}, 1255-1263 (2007).

\bibitem{jean1}  R. V. Jean, ``Nomothetical modelling of spiral symmetry in biology''  in ``Five fold symmetry'', editor I. Hargittai, \emph{World Scientific, Singapore}, $505-528$ (1992).

\bibitem{jean2}  R. V. Jean,  ``Mathematical Modeling in Phyllotaxis: The State of the Art''  \emph{Mathematical Biosciences} \textbf{64}, $1-27$ (1983).


\bibitem{ridley} I. N. Ridley, ``Packing efficiency in sunflower heads''
\emph{Mathematical Biosciences} \textbf{58}, $129-139$ (1982).

\bibitem{orgel} J. Orgel ,  A. Miller, T.C. Irwing , R.F. Fischetti , A.P. Hammersley  and  T.J. Wess, ``The in situ
supermolecular structure of type I collagen.'' \emph{ Structure}   \textbf{9}, $1061-1069$ (2001).


\bibitem{rivier84} N. Rivier, R. Occelli,  J. Pantaloni  and  A. Lissowski, ``Structure of B\'enard convection cells,
phyllotaxis and crystallography in cylindrical symmetry'' \emph{ J. Phys, France} \textbf{45}, $49-63$ (1984).

\bibitem{rivier92}  N. Rivier, ``The structure and dynamics of patterns of
Bernard convection cells'', \textit{J. Phys.:Condens. Matter} \textbf{4}, $913-943$ (1992).


\bibitem{jelinsky} L. W. Jelinsky, C. E. Sullivan and D. A. Torchia, ``H NMR study of molecular motion in collagen fibrils''  \textit{Nature} \textbf{284}, $531-534$ (1980).


\end{thebibliography}
\end{document}